\documentclass[twocolumn]{jpsj2}

\title{
  Population Coding by Globally Coupled Phase Oscillators }

\author{
  Hiroya Nakao\thanks{nakao@ton.scphys.kyoto-u.ac.jp} }

\inst{
  Department of Physics, Kyoto University, Kyoto 606-8502 }

\abst{
  A system of globally coupled phase oscillators subject to an
  external input is considered as a simple model of neural circuits
  coding external stimulus.
  The information coding efficiency of the system in its asynchronous
  state is quantified using Fisher information.
  The effect of coupling and noise on the information coding
  efficiency in the stationary state is analyzed.
  The relaxation process of the system after the presentation of an
  external input is also studied.
  It is found that the information coding efficiency exhibits a large
  transient increase before the system relaxes to the final stationary
  state.
}

\recdate{\today}

\kword{coupled oscillators, neural networks, information coding, statistical inference}

\begin{document}

\maketitle

\section{Introduction}

Oscillatory activity of neurons is ubiquitous in various areas of the
brain at various scales, whose physiological relevance to the
information processing has been discussed in a number of
studies~\cite{Gray,Engel,NatureReviews1,NatureReviews2}.
In modeling cortical neural circuits, coupled oscillators have played
important roles~\cite{Golomb,Ermentrout}. In this framework, the
neurons are modeled as mutually interacting limit-cycle oscillators,
where simplified interaction rules between the oscillators are often
assumed for the sake of mathematical tractability.
The case of global (all-to-all) coupling is typical of such simplified
interactions, where the oscillators interact through the mean field of
all oscillators. The prominent feature of globally coupled oscillators
is complete synchronization~\cite{Golomb,Kuramoto,Kuramoto2,Pikovsky}.
However, it is experimentally known that the cortical neurons in vivo
rarely exhibit large-scale complete synchronization, but usually
exhibit irregular firing patterns~\cite{Softky}. Therefore, it has
been discussed how globally coupled models of neural populations can
sustain asynchronous firing
activity~\cite{Abbott,Gerstner,Hansel,Mar,Omurtag,Nykamp}.

It is generally considered that the cortical information processing is
achieved through population coding, namely, collective representation
of information by large numbers of
neurons~\cite{Paradiso,Seung,NeuralCodes}. Regarding quantification of
information coding efficiency by a population of neurons, a framework
based on statistical estimation theory has been utilized.
In this framework, the information coding efficiency is quantified by
how precisely the given stimulus can be estimated from the observed
firing rates of those neurons. It can be measured by the Fisher
information, which gives the accuracy of parameter estimation in
statistical estimation theory~\cite{Lehman,Cover}.
Using this framework, the information coding efficiency of various
stochastic neuron models has been
calculated~\cite{Paradiso,Seung,NeuralCodes}.

In this paper, instead of stochastic neuron models, we introduce a
system of globally coupled phase oscillators as a simple model of
cortical neural circuits. Our system exhibits asynchronous state where
each oscillator rotates (``fires'') independently, by which we model
the ``irregular firing'' of the cortical neurons.
It codes the parameter value of an external input in the asynchronous
phase distribution of the oscillators. We quantify the information
coding efficiency of our system using the statistical estimation
framework. We discuss the effect of coupling and noise on the
information coding efficiency in the asynchronous stationary state.
We also study the dynamical aspect of our system. Even if the
oscillators are not in strict synchrony with each other, their
population density can still exhibit coherent damped oscillation. We
show that such oscillation can lead to a large transient increase in
the information coding efficiency.

\begin{figure}[htbp]
  \begin{center}
    \includegraphics[width=0.4\textwidth, clip=true]{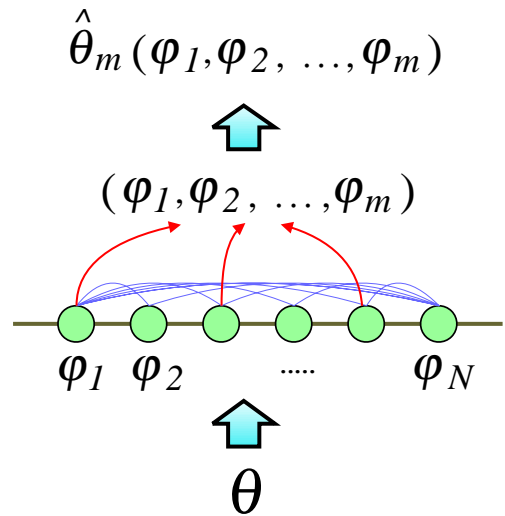}
  \end{center}
  \caption{ Population coding by globally coupled phase oscillators.
    Information coding efficiency of the system is quantified using
    Fisher information, which gives the accuracy of an estimator
    $\hat{\theta}_{m}(\varphi_1, ..., \varphi_m)$ of the external
    input parameter $\theta$ from observed phase variables $\varphi$
    of the oscillators. }
  \label{fig:01}
\end{figure}

\section{Globally coupled phase oscillators}

We use a system of globally coupled phase oscillators, which was
previously studied in detail by Golomb {\it et al.}~\cite{Golomb}.  It
exhibits asynchronous states as well as various synchronized states.
We set the system at its asynchronous state, and add a constant
external input to the system that depends on a given external
parameter $\theta$. This parameter $\theta$ is then coded by the
non-uniform phase distribution of the oscillators (Fig.~\ref{fig:01}).
Our interest is how efficiently the population of the oscillators
codes this external parameter. We quantify it by how precisely we can
estimate the given input by observing the phase variables of the
oscillators.
This system may be considered, for example, as a simple qualitative
model of the orientation column in the visual
cortex~\cite{Paradiso,Seung,Omurtag,Nykamp}. In that case, the
external parameter $\theta$ corresponds to the angle of a visual
directional stimulus.

\begin{figure}[htbp]
  \begin{center}
    \includegraphics[width=0.4\textwidth, clip=true]{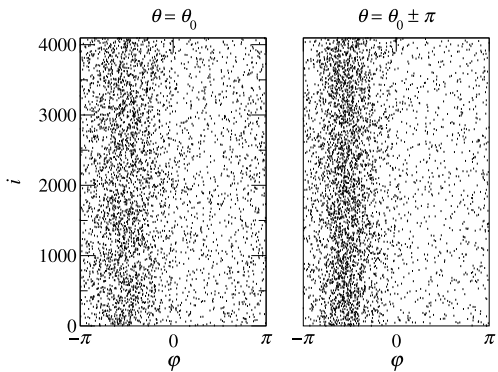}
  \end{center}
  \caption{ Snapshots of the phase $\varphi$ of the oscillators at two
    values of the external input, $\theta = \theta_0$ and $\theta =
    \theta_0 \pm \pi$.}
  \label{fig:02}
\end{figure}

Our system consists of $N$ identical phase oscillators mutually
interacting through their mean field, which obey the following
Langevin equations:
\begin{equation}
  \frac{d}{dt} \varphi_{i}(t) = f(\varphi_{i}(t))
  + \frac{1}{N} \sum_{j=1}^{N} g(\varphi_{j}(t)) + \xi_{i}(t)
  + H(\theta)
  \label{Eq:Langevin}
\end{equation}
for $i = 1, ..., N$, where $\varphi_{i}$ represents the phase of the
$i$-th oscillator, $f(\varphi) = A + \sin \varphi$ the individual
dynamics of an isolated oscillator, $g(\varphi) = C \sin(\varphi +
\alpha)$ the interaction between the oscillators, and $\xi_{i}$ a
Gaussian-white noise with zero mean, whose correlation function is
given by $ \langle \xi_i(t) \xi_j (t') \rangle = 2 D \delta (t-t')
\delta_{i, j} $.
We identify $\varphi + 2 \pi k$ ($k$ is an integer number) with
$\varphi$ and restrict the value of $\varphi$ in $[-\pi, \pi]$.
$H(\theta)$ is a $\theta$-dependent constant external input common to
all oscillators, which we explain below.
The parameter $A$ determines the dynamics of each oscillator, $C$ the
coupling strength, $\alpha$ the phase shift of coupling, and $D$ the
intensity of noise.
We assume that $C$ can take both positive and negative values, and the
range of $\alpha$ to be $(0, \pi)$. We exclude two extreme values
$\alpha = 0$ and $\alpha = \pi$, since the system becomes singular
with these values~\cite{Golomb}.

When isolated, each oscillator exhibits both excitatory and
self-oscillatory dynamics depending on the value of $A$ (without the
noise and the external input, each oscillator is self-oscillatory when
$A>1$).
To realize the asynchronous state, we set $A$ at a sufficiently large
value, so that each oscillator is self-oscillatory even if the effects
of the coupling, the noise, and the external input are incorporated.
In this self-oscillatory situation, the mean rotation rate (or the
``firing rate'') of the oscillator increases with $A$.

Through the external input $H(\theta)$, the dynamics of the phase
$\varphi$ is affected by the external parameter $\theta$.
We assume the range of $\theta$ to be $[-\pi, \pi]$ (``angle
stimulus'') and use a functional form $H(\theta) = H_0 \cos( \theta -
\theta_{0} )$ as the external input, where $H_0$ determines its
strength and $\theta_0$ the location of its maximum (``preferred
stimulus'').
This $H(\theta)$ roughly models the one-humped ``tuning curve'', which
represents the response of a neuron to the
stimulus~\cite{Paradiso,Seung,NeuralCodes}.
When $\theta$ approaches $\theta_0$, the mean rotation rate of the
oscillators increases, and reaches the maximum value at $\theta =
\theta_{0}$ (in the following discussion, the results depends only on
the difference $\theta - \theta_0$, so that the absolute value of
$\theta_0$ is not important).
Since this external input $H(\theta)$ merely shifts the parameter $A$
in the individual dynamics of the oscillator $f(\varphi) = A + \sin
\varphi$, we hereafter include $H(\theta)$ in $A$ and denote the
effective value of $A$ as $ A(\theta) = A + H(\theta) $.

Figure~\ref{fig:02} displays two snapshots of the phase $\varphi_i$ of
all oscillators in the asynchronous stationary state obtained by
direct numerical simulations of the Langevin
equations~(\ref{Eq:Langevin}) at two different values of the external
input, $\theta = \theta_0$ and $\theta = \theta_0 \pm \pi$ ($\theta =
\theta_0 + \pi$ and $\theta_0 - \pi$ give the same external input
because $H(\theta)$ is periodic in $\theta$). The other parameters are
fixed at $A=1.5$, $C=0.5$, $\alpha=\pi/4$, $H_0=0.1$, and $D=0.1$.  It
can be seen that the distribution of the phase is relatively uniform
at the preferred input $\theta = \theta_0$, whereas its non-uniformity
is enhanced at $\theta = \theta_0 \pm \pi$.

Hereafter, rather than tracking the individual stochastic trajectories
of the oscillators directly, we take the $N \to \infty$ limit and
consider the one-body probability density function (PDF) $P(\varphi, t
; \theta)$ of the phase $\varphi$.
The evolution of the PDF $P(\varphi, t ; \theta)$ is described
by~\cite{Golomb,Kuramoto,Kuramoto2}
\begin{eqnarray}
  \label{Eq:PDF_evol}
  \frac{\partial}{\partial t} P(\varphi, t ; \theta) &=&
  -\frac{\partial}{\partial \varphi} I(\varphi, t ; \theta), \cr \cr
  I(\varphi, t ; \theta) &=&
  \left\{ A(\theta)
    + G(t ; \theta) + \sin \varphi \right\} P(\varphi, t ; \theta) \cr \cr
  &&- D \frac{\partial}{\partial \varphi}  P(\varphi, t ; \theta),
\end{eqnarray}
where $I(\varphi, t ; \theta)$ represents the probability flux, and
$G(t ; \theta)$ the ``internal field'' defined as the interaction
function averaged by the PDF $P(\varphi, t ; \theta)$, i.e.,
\begin{equation}
  \label{Eq:self_consistent_g}
  G(t ; \theta) = \int_{- \pi}^{\pi} P(\varphi, t ; \theta)
  C \sin(\varphi + \alpha) d\varphi.
\end{equation}
Periodic boundary conditions are assumed for $P(\varphi, t ; \theta)$
and $I(\varphi, t ; \theta)$ at $\varphi = \pm \pi$.

\begin{figure}[htbp]
  \begin{center}
    \includegraphics[width=0.4\textwidth, clip=true]{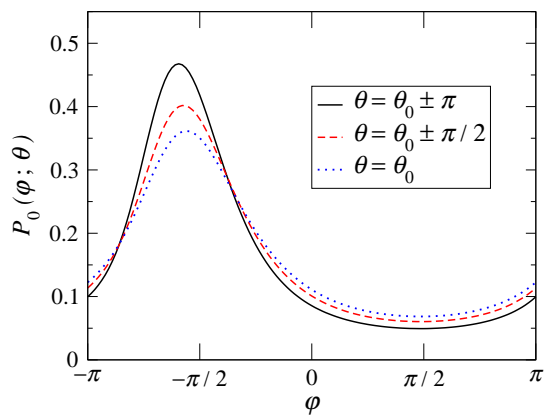}
  \end{center}
  \caption{ Stationary PDF $P_{0}(\varphi ; \theta)$ at several values
    of the external parameter $\theta$ ($\theta = \theta_0 \pm \pi$,
    $\theta = \theta_0 \pm \pi/2$, and $\theta = \theta_0$).}
  \label{fig:03}
\end{figure}

Figure~\ref{fig:03} displays stationary PDFs of the phase $P_0(\varphi
; \theta)$ at several values of the external parameter $\theta$ in the
asynchronous stationary state, which corresponds to Fig.~\ref{fig:02}.
The details of this state will be given in the next section.
The PDF is relatively uniform at the preferred input $\theta =
\theta_0$, and becomes steeper at $\theta = \theta_0 \pm \pi$.
At this weak noise level, $D=0.1$, functional shape of the PDF is
close to that in the noiseless case.
As the value of $\theta$ is varied, the shape of $P_0(\varphi ;
\theta)$ varies correspondingly. In other words, the external
parameter $\theta$ is coded by the stationary PDF in the asynchronous
state of our globally coupled phase oscillators.

\section{Asynchronous stationary state}

In this section, we summarize the dynamics of Eqs.~(\ref{Eq:PDF_evol})
and (\ref{Eq:self_consistent_g}) following Golomb {\it et
  al.}~\cite{Golomb} with emphasis on the asynchronous stationary
state.
The PDF $P(\varphi,t ; \theta)$ exhibits mainly three behavior,
namely, fixed state, limit-cycle state, and asynchronous stationary
state.
The probability flux $I(\varphi, t ; \theta)$ takes a small value in
the fixed state (zero in the noiseless case), a non-zero constant
value in the asynchronous stationary state, and oscillates
periodically in the limit-cycle state.

Let us consider the noiseless case ($D=0$) first. In this case,
existence and linear stability of each state can be studied
analytically~\cite{Golomb}.
Here, it should be noted that even if we set $D=0$, we always assume
that the system is subject to infinitesimally weak noise, to avoid
singular behavior specific to the noiseless equation and to ensure the
existence of final steady states~\cite{Kuramoto2}.
With this understanding, we formally set $D=0$ to facilitate
analytical treatment, and compare the analytical results with the
numerical simulations at finite values of $D$.

The fixed-state solutions exist when
$C^2 + 2 C \cos\alpha + 1 > A(\theta)^2$, and the stability conditions
are given by $C \cos\alpha > -1+A(\theta)$ or by $C \cos \alpha > -1 -
A(\theta)$ and $C > 0$. In this state, all oscillators take the same
fixed phase homogeneously, and the PDF is simply a $\delta$-function
with fixed location. The probability flux is constantly zero.
The limit-cycle solution exists when
$C^2 + 2 C \cos\alpha + 1 < A(\theta)^2$, and is stable when $C > 0$.
In this state, the oscillators exhibit completely synchronized
rotation. The PDF is again a $\delta$-function, whose location rotates
steadily. The temporal sequence of the probability flux is given by
periodically aligned $\delta$-functions.

In our context, these two states are inappropriate for modeling the
dynamics of cortical neurons. In the fixed state, all oscillators are
trapped to the same fixed phase and never rotate, namely, the neurons
do not fire at all. On the other hand, the limit-cycle state
corresponds to the completely synchronized firing of all neurons,
which is experimentally unrealistic. These situations are not changed
even if weak noise is introduced. Only with sufficiently strong noise,
these states may be utilized in modeling the dynamics of cortical
neurons, but we do not consider such cases in this paper for
simplicity.

The asynchronous stationary solution exists when $-\infty < C
\cos\alpha < [ A(\theta)^2 - 1 ] / 2$.  It coexists with the
limit-cycle or fixed solution.
Unlike the above two cases, the PDF takes a broad functional form in
this state.
The oscillators are not in synchrony with each other, and rotate with
a finite constant rate steadily.
The probability flux $I(\phi, t ; \theta)$ takes a constant value,
which is typically between $0.05$ and $0.20$ at the parameter values
we use in this paper.
The stationary PDF $P_{0}(\varphi ; \theta)$ in the asynchronous
stationary state at $D=0$ can be obtained from
Eqs.~(\ref{Eq:PDF_evol}) and (\ref{Eq:self_consistent_g}) as
\begin{equation}
  \label{Eq:PDF_stationary}
  P_{0}(\varphi ; \theta) =
  \frac{ \left[ F(\theta)^2 - 1 \right]^\frac{1}{2} }{2 \pi}
  \frac{1}{F(\theta) + \sin \varphi}.
\end{equation}
Here, $F(\theta)$ is defined as $ F(\theta) = A(\theta) +
G_{0}(\theta) $, where $G_{0}(\theta)$ denotes the internal field in
this stationary state. The value of $G_{0}(\theta)$, or equivalently
the value of $F(\theta)$, is determined self-consistently so that
$G_{0}(\theta)$ and $P_{0}(\varphi ; \theta)$ satisfy
Eq.~(\ref{Eq:self_consistent_g}). It is explicitly calculated as
\begin{eqnarray}
  F(\theta)
  &=& \frac{1}{1 + 2 C \cos \alpha}
  \{ (1 + C \cos \alpha) A(\theta)
  \cr \cr
  &\pm&
  \sqrt{(C \cos \alpha)^2 [ A(\theta)^2 - (1 + 2 C \cos \alpha) ]}
  \}.
  \label{Eq:self_consistent_F}
\end{eqnarray}
When $C=0$ or $\alpha = \pi/2$ ($C \cos \alpha=0$),
we obtain a trivial solution $F(\theta) = A(\theta)$, i.e.,
$G_{0}(\theta) = 0$.
The relevant solution for $C \cos \alpha \neq 0$ is given by the
negative branch when $-\infty < C \cos \alpha < 0$, and by the
positive branch when $0 < C \cos \alpha < [ A(\theta)^2 - 1 ]/2$.
When $C \cos \alpha > [A(\theta)^2 - 1]/2$, no stationary PDF exists
that corresponds to the asynchronous stationary state.

\begin{figure}[htbp]
  \begin{center}
    \includegraphics[width=0.4\textwidth, clip=true]{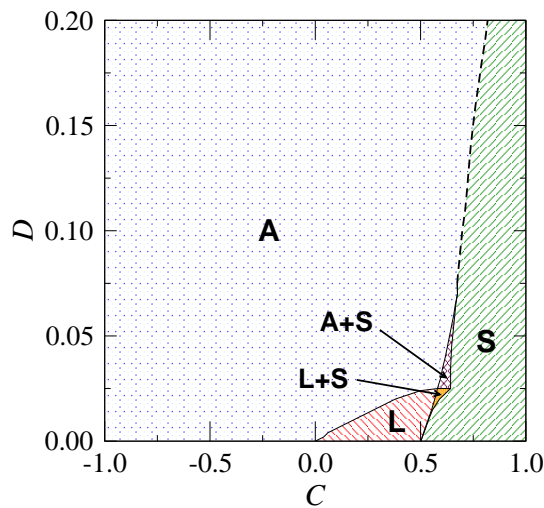}
  \end{center}
  \caption{ Phase diagram of the system at $A = 1.4$. Region ``A''
    represents the asynchronous stationary state, ``L'' the
    limit-cycle state, and ``S'' the fixed state. In the small region
    ``A+S'', the asynchronous state and the fixed state are bistable,
    and in the region ``L+S'', the limit-cycle state and the fixed
    state are bistable. The dotted curve distinguishes the regions
    ``A'' and ``S'' as defined in the text. }
  \label{fig:04}
\end{figure}

For example, when $0 < \alpha < \pi/2$, the non-uniformity of
$P_0(\varphi ; \theta)$ is enhanced as we increase $C$, because
$F(\theta)$ decreases.
However, as we increase $C$ further, $C \cos \alpha$ eventually
reaches $[ A(\theta)^2 - 1 ] / 2$, and the stationary PDF
corresponding to the asynchronous stationary state ceases to exist.
We are then left with a $\delta$-peaked stationary PDF that
corresponds to the fixed state of the oscillators.

According to the linear stability analysis of the stationary
PDF~\cite{Golomb}, $P_0(\varphi ; \theta)$ is marginally stable when
$C < 0$, and is unstable when $C > 0$ in the absence of noise.
Therefore, when $C > 0$, the asynchronous stationary state cannot be
realized without noise, and only the fixed or limit-cycle state is
observed in numerical simulations.

When the noise exists ($D>0$), the functional form of the stationary
PDF becomes complex. However, we can still obtain it numerically by
finding the stationary solution of Eqs.~(\ref{Eq:PDF_evol}) and
(\ref{Eq:self_consistent_g}).
The $\delta$-peaked PDF in the fixed or limit-cycle state at $D=0$ is
smeared by the noise, and takes broader functional form. The
asynchronous stationary state is also broadened by the noise.
Furthermore, the noise also stabilizes the asynchronous stationary
state~\cite{Golomb}, so that originally only marginally stable PDF at
$C<0$ becomes linearly stable whenever $D>0$, and even the originally
unstable asynchronous stationary state at $C>0$ can be stabilized when
$D>D_c$, where $D_c$ is a certain critical noise intensity that
depends on $C$.

Figure~\ref{fig:04} displays a numerically obtained phase diagram of
our system, where the final steady state of the system sufficiently
after the initial transient is indicated as a function of the coupling
strength $C$ and the noise intensity $D$.
The parameter $A(\theta)$ is fixed to its minimum value, $A(\theta =
\theta_0 \pm \pi) = A - H_0 = 1.4$, because the stability region of
the asynchronous stationary state, which is of our primary interest,
becomes smallest at this value.
As can be seen, originally unstable asynchronous stationary state for
$C>0$ can easily be stabilized as $D$ becomes larger than a certain
small critical value $D_c$.

In region ``A'' of the phase diagram, the PDF exhibits stable
asynchronous stationary state, which continues from the marginally
stable ($C<0$) or unstable ($C>0$) asynchronous solution at $D=0$.
In region ``L'', the PDF exhibits limit-cycle oscillation, which is
the continuation of the completely synchronous limit-cycle solution at
$D=0$.
In region ``S'', the PDF exhibits fixed stationary state with small
probability flux, which continues from the homogeneous fixed solution
at $D=0$.
There also exist small bistable regions where the fixed state coexists
with the asynchronous stationary state or with the limit-cycle state
at small noise intensity, roughly $D < 0.07$, as indicated in the
figure as ``A+S'' or ``L+S''.
As $D$ becomes larger than $0.07$, the bistable region vanishes, and
the distinction between the asynchronous stationary state and the
fixed state becomes unclear.
We therefore distinguish the asynchronous stationary state ``A'' from
the fixed state ``S'' by whether the probability flux $I(\varphi, t ;
\theta)$ exceeds $0.05$ or not when $D > 0.07$.
Here, the value $0.05$ is determined using the probability flux at the
upper endpoint of the bistable region ``A+S''. 
The boundary between ``A'' and ``S'' determined in this way is
displayed in the figure by the dotted curve.
As we cross this boundary from region ``A'' to ``S'', the PDF changes
its shape and the probability flux drops quickly.
Hereafter, we restrict our analysis to the asynchronous stationary
state ``A'' with sufficiently large probability flux. The fixed state,
the limit-cycle state, and the bistable regions are excluded from our
consideration.

\section{Fisher information}

Let us quantify the information coding efficiency of our system using
Fisher information.
We consider the accuracy of the parameter estimation of $\theta$ from
$m$ samples of the phase $\varphi_{1}, ..., \varphi_{m}$ observed
independently from the $\theta$-dependent PDF $P(\varphi ; \theta)$.
According to the statistical estimation theory~\cite{Lehman,Cover},
the maximum likelihood estimator $\hat{\theta}_{m}(\varphi_{1}, ...,
\varphi_{m})$, which is determined so as to maximize the likelihood
function $ L(\theta ; \varphi_{1}, ..., \varphi_{m}) = \prod_{k=1}^{m}
P(\varphi_{k} ; \theta) $, weakly converges to a normal distribution
with mean $\theta$ and variance $1 / m J(\theta)$ in the limit of
large $m$.
The quantity $J(\theta)$ in the denominator of the variance is the
Fisher information, which is defined as
\begin{equation}
  \label{Eq:Fisher_def}
  J(\theta)
  = \int_{- \pi}^{\pi} P(\varphi ; \theta)
  \left[ \frac{\partial}{\partial \theta} 
    \log P(\varphi ; \theta)
  \right]^2 d\varphi.
\end{equation}
Under some regularity conditions, the maximum likelihood estimator is
proven to be asymptotically optimal in the large $m$ limit, in the
sense that asymptotic variance of all other asymptotically normal
estimators cannot be smaller than $1 / m
J(\theta)$~\cite{Lehman,Cover}.
Thus, the Fisher information $J(\theta)$ gives a lower bound to the
error of parameter estimation, and it can be considered as quantifying
the information (actually parameter) coding efficiency of the system
by its PDF.
As can be seen from Eq.~(\ref{Eq:Fisher_def}), $J(\theta)$ measures
the mean ``response'' of the PDF $P(\varphi ; \theta)$ to a slight
change in the parameter $\theta$. When this response is large, the
estimation error becomes small, and the information coding efficiency
becomes large.
The Fisher information can also be related to the discriminability
measure between two similar stimuli used in
psychophysics~\cite{Paradiso,Seung}.

\begin{figure}[htbp]
  \begin{center}
    \includegraphics[width=0.4\textwidth, clip=true]{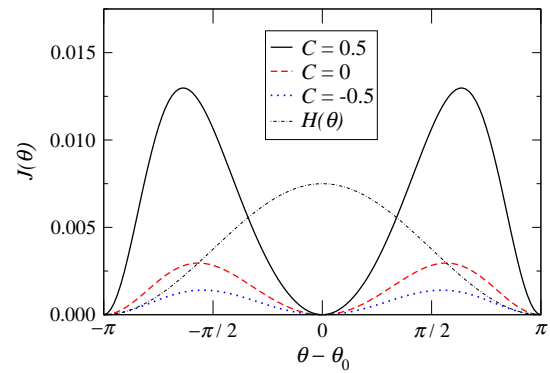}
  \end{center}
  \caption{ Fisher information $J(\theta)$ obtained for the uncoupled
    case ($C=0$) and for the coupled cases ($C = \pm 0.5$). Cosine
    curve in the lower part is the external input (``tuning curve'')
    $H(\theta)$ given to the system (rescaled and shifted).}
  \label{fig:05}
\end{figure}

\section{Stationary information coding efficiency}

We first consider the information coding efficiency in the stationary
state of our system. In the absence of noise ($D=0$), we can
analytically calculate the Fisher information $J(\theta)$ from the
self-consistent stationary PDF $P_0(\varphi ; \theta)$,
Eq.~(\ref{Eq:PDF_stationary}), as
\begin{equation}
  J(\theta)
  = \frac{1}{2} \left[ \frac{\partial F(\theta)}{\partial \theta} \right]^2
  \frac{1}{\left[ F(\theta)^2 - 1 \right]^2}.
\end{equation}
Since $F(\theta)$ contains the self-consistent internal field
$G_{0}(\theta)$, $J(\theta)$ depends on the mutual coupling through
this quantity.
By substituting the explicit form of $F(\theta)$,
Eq.~(\ref{Eq:self_consistent_F}), we can express $J(\theta)$ in terms
of $A(\theta) = A + H(\theta)$ and $C \cos \alpha$.
When $D>0$, we can calculate $J(\theta)$ using numerically obtained
stationary PDFs.
With the weak noise we use in this paper, the numerically obtained
Fisher information for $D>0$ is close to the noiseless theoretical
values at $D=0$.

Figure~\ref{fig:05} displays the Fisher information $J(\theta)$
obtained for the uncoupled case ($C = 0$) and for the coupled cases
($C = \pm 0.5$) as a function of the external parameter $\theta$. The
other parameters are fixed at $A=1.5$, $\alpha = \pi / 4$, $H_0 =
0.1$, and $D=0.1$.
The Fisher information clearly depends on the coupling strength $C$;
we attain much larger value at $C = 0.5$ than at $C=0$ or $C=-0.5$.
At this noise level, the numerically obtained Fisher information is
close to the noiseless theoretical values.
Note here that the asynchronous stationary state is unstable without
the noise when $C > 0$, so that the corresponding theoretical values
of the Fisher information cannot be attained in practice.
However, as can be seen from Fig.~\ref{fig:04}, the weak noise
($D=0.1$) can readily stabilize the originally unstable PDF at $C > 0$
and helps realize much larger information coding efficiency.

\begin{figure}[htbp]
  \begin{center}
    \includegraphics[width=0.4\textwidth, clip=true]{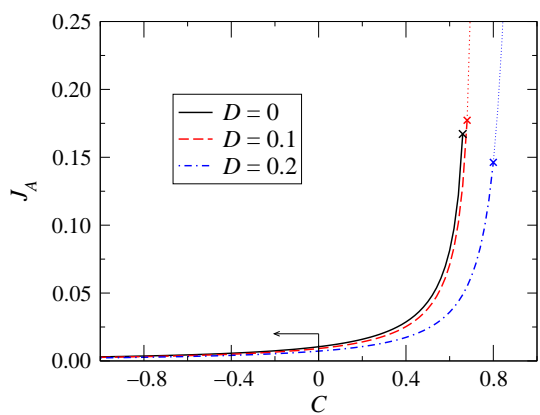}
  \end{center}
  \caption{ Dependence of the mean Fisher information $J_A$ on the
    coupling strength $C$. At $D=0$, only the region indicated by the
    arrow can be realized. Crosses represent the values of $C$ where
    the asynchronous stationary state disappears ($D=0$) or gives way
    to the stationary fixed state ($D=0.1$, $D=0.2$).}
  \label{fig:06}
\end{figure}

Since our external input $H(\theta)$ takes its maximum value at
$\theta = \theta_{0}$, the mean rotation rate of each oscillator is
also maximized at this value.
However, $J(\theta)$ vanishes at this point, because $\theta =
\theta_{0}$ is the extremum of $A(\theta)$, hence $\partial F /
\partial \theta \propto \partial A / \partial \theta = 0$ holds.
Thus, accurate parameter estimation is difficult around $\theta =
\theta_{0}$. Rather, $J(\theta)$ takes its maximum at $\theta \simeq
\pm \pi / 2$, where the functional shape of the stationary PDF most
strongly depends on the external parameter $\theta$.
This fact is consistent with the results in the previous
studies~\cite{Paradiso,Seung,NeuralCodes}, which is interesting since it is
against our rate-coding intuition.

In Fig.~\ref{fig:06}, the dependence of the mean Fisher information
$J_A = (2 \pi)^{-1} \int_{- \pi}^{\pi} J(\theta) d\theta$ on the
coupling strength $C$ is displayed at three values of the noise
intensity, $D=0$, $D=0.1$, and $D=0.2$.
When $D=0$, only the region $C < 0$ as indicated in the figure is
realizable. However, when $D=0.1$ or $0.2$, originally unstable PDF
for $C > 0$ is stabilized as shown in Fig.~\ref{fig:04}, so that the
whole curve can be realized.
$J_A$ increases remarkably as $C$ is increased, because the
non-uniformity of the stationary PDF is enhanced, and its dependence
on the parameter $\theta$ becomes stronger.

\begin{figure}[htbp]
  \begin{center}
    \includegraphics[width=0.4\textwidth, clip=true]{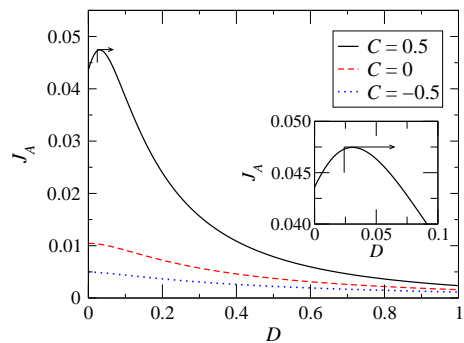}
  \end{center}
  \caption{ Dependence of the mean Fisher information $J_A$ on the
    noise intensity $D$. At $D=0$, only the region indicated by the
    arrow can be realized.}
  \label{fig:07}
\end{figure}

If $C$ is further increased to a certain critical value, the
stationary PDF corresponding to the asynchronous stationary state
disappears ($D=0$) or gives way to the fixed state with low
probability flux ($D=0.1$, $D=0.2$). These values are indicated by
crosses in the figure.
The Fisher information diverges at this point when $D=0$, because the
stationary PDF changes its functional form discontinuously.
When $D=0.1$ and $D=0.2$, divergence of the Fisher information is
suppressed, but it still exhibits abrupt increase due to the sudden
change in the stationary PDF.
Although this result is formally valid, such divergent behavior
indicates the failure of our present framework based on the Fisher
information in quantifying the information coding efficiency around
bifurcation points; it implies that we can estimate the external
parameter with infinite precision even from a single observation.
This is due to the basic assumption of our framework, namely,
continuous dependence of the functional shape of the PDF on the
external parameter, which is generally violated at the bifurcation
points.
Thus, at present, we consider the divergent behavior of the Fisher
information around the bifurcation point as rather a mathematical
artifact.
In conventional studies of the information coding efficiency using
stochastic neuron models~\cite{Paradiso,Seung,NeuralCodes}, the firing
statistics of the neurons is usually assumed to be given by some
specific family of the PDF, e.g. Gaussian or Poissonian, so that such
a situation does not arise generally.

Figure \ref{fig:07} displays the dependence of the mean Fisher
information $J_A$ on the noise intensity $D$ at three values of the
coupling strength, $C=-0.5$, $C=0$, and $C=0.5$. The other parameters
are fixed at $A=1.5$, $\alpha = \pi / 4$, and $H_0 = 0.1$.
As previous, the asynchronous stationary state is unstable at small
$D$ when $C=0.5$, and only the region indicated by the arrow in the
figure can be realized.
As we increase $D$, $J_A$ decreases. The noise generally decreases the
information coding efficiency of the system, because the noise tends
to flatten the stationary PDF and weakens its dependence on $\theta$.
However, interestingly, the mean Fisher information $J_A$
corresponding to $C=0.5$ takes its maximum value not at $D=0$ but at a
small but finite value of $D$, as shown in the inset of
Fig.~\ref{fig:07} (the asynchronous stationary state is already
stabilized at this value of $D$).
Thus, the noise is generally disturbing, but it can also help the
realization of larger information coding efficiency.

\begin{figure}[htbp]
  \begin{center}
    \includegraphics[width=0.4\textwidth, clip=true]{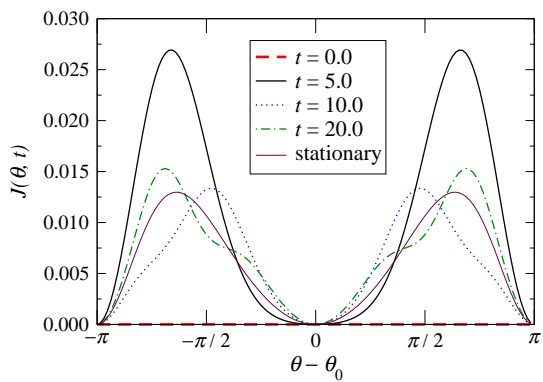}
  \end{center}
  \caption{ Snapshots of Fisher information $J(\theta, t)$ at several
    moments in the relaxation process after the onset of the external
    input.  }
  \label{fig:08}
\end{figure}

\section{Dynamic information coding efficiency}

Since our system is a stochastic dynamical system, we can consider not
only its stationary state, but also its non-stationary state. We here
briefly study the relaxation process after the presentation of an
external input.
Let us assume that the system is in its initial asynchronous
stationary state without any external input when $t<0$. After $t=0$,
we apply a constant external input $H(\theta)$ to the system and
observe its relaxation to a new asynchronous stationary state
determined by the given parameter $\theta$.
By solving the evolution equation (\ref{Eq:PDF_evol}) of the PDF
$P(\varphi, t ; \theta)$ numerically, we can calculate the Fisher
information $J(\theta, t)$ at any moment.

\begin{figure}[htbp]
  \begin{center}
    \includegraphics[width=0.4\textwidth, clip=true]{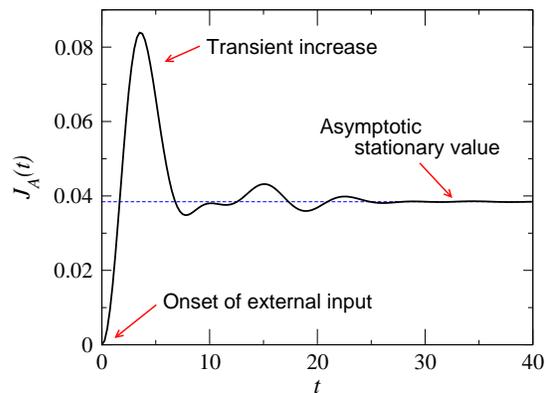}
  \end{center}
  \caption{ Temporal variation of mean Fisher information $J_A(t)$
    after the onset of the external input.}
  \label{fig:09}
\end{figure}

Figure \ref{fig:08} shows several snapshots of $J(\theta, t)$ during
the relaxation process, and Figure~\ref{fig:09} shows temporal
variation of the mean Fisher information $J_A(t) = (2 \pi)^{-1}
\int_{- \pi}^{+ \pi} J(\theta, t) d\theta$. The system parameters are
$A = 1.5$, $\alpha = \pi / 4$, $C = 0.5$, $H_0 = 0.1$, and $D = 0.1$.
Interestingly, slightly after the onset of the external input,
$J(\theta, t)$ exhibits wavy functional shape that is considerably
different from the previously obtained stationary functional shape,
and then it gradually converges to the stationary shape.
Correspondingly, after the onset of the external input, $J_A(t)$
rapidly increases from zero to a certain maximum value and then
exhibits an oscillatory relaxation to the new stationary value.

This reflects the fact that the dependence of the transient PDF
$P(\varphi, t ; \theta)$ on $\theta$ is much stronger than that of the
stationary PDF $P_{0}(\varphi ; \theta)$, which can be qualitatively
understood as follows. When we apply a constant external input
$H(\theta)$, the PDF undergoes oscillatory relaxation to the new
stationary state, with the period of the oscillation determined by the
parameter $\theta$.
In this process, the discrepancy of two PDFs corresponding to two
slightly different values of $\theta$ becomes much larger than that in
the stationary state, through a mechanism essentially similar to the
``beat'' of sound waves.
As a result, the Fisher information, which measures the mean response
of the PDF to a slight change in the parameter, also increases
remarkably and exhibits a transient increase in information coding
efficiency.

\section{Summary}

We considered population coding of an external parameter in the
asynchronous state of globally coupled phase oscillators and
quantified its information coding efficiency using Fisher information.
The Fisher information depends on the mutual coupling through the
self-consistent internal field of the oscillators.
The noise generally reduces the Fisher information of the system, but
it also stabilizes the originally unstable PDF and helps realize
larger information coding efficiency.
We found a substantial transient increase in Fisher information
slightly after the onset of an external input, before the system
relaxes to a new stationary state.

Although we treated the information coding only in the asynchronous
state in this paper, the fixed or limit-cycle state with sufficiently
strong noise may also be considered as the basis for the information
coding.
It would then be interesting to compare the information coding
efficiency among those dynamically distinct situations.
However, as we saw in Fig.~\ref{fig:06}, our framework based on Fisher
information does not yield practically meaningful result around the
bifurcation point between different dynamical states.
To remedy this situation, further modifications to our present
framework will be necessary.

The transient increase in Fisher information indicates that the
extraction of the information stored in our system can be better
achieved in the transient state than in the final stationary state.
Though our present system is only a rough caricature of real cortical
networks, the mechanism leading such behavior can be generic. Thus, it
could provide an interesting viewpoint on the role played by the
transient dynamical behavior of the population of real cortical
neurons, e.g. in interpreting the results of psychophysical
experiments.

In any case, to assert the relevance of our physical findings in the
actual information processing by the cortical neurons, we need careful
discussions on the physiological plausibility of the model.
A more detailed analysis of the present system and generalization to
more realistic neural models will be reported in the future.

Finally, from a general viewpoint, our analysis reported in this paper
can be considered as a prototypical study of physical systems coding
external information. Such a framework would be important in analyzing
physical systems that process external information, not only in the
context of neuroscience but also in various other areas.

\acknowledgments The author thanks Y. Sakai, M. Hayashi, Y. Tsubo, and
H.  Shimazaki for useful comments and suggestions. He is also grateful
to S. Amari, M. Okada, and the members of the RIKEN Brain Science
Institute for their advice and support.

\end{document}